\documentstyle[aps,twocolumn,epsfig,floats]{revtex}

\begin{document}  
\draft
\wideabs{  
\author{Christof Gattringer}
\address{\medskip Institut f\"ur Theoretische Physik, Universit\"at
Regensburg, D-93040 Regensburg, Germany} 
\date{February 7, 2002}
\title{Testing the self-duality of topological lumps in SU(3) 
lattice gauge theory}
\maketitle
   
\begin{abstract}  
We discuss a simple formula which connects the field-strength tensor
to a spectral sum over certain quadratic forms of the eigenvectors
of the lattice Dirac operator. We analyze these terms for the near 
zero-modes and find that they give rise to contributions which are 
essentially either
self-dual or anti self-dual. Modes with larger eigenvalues in the bulk of the 
spectrum are more dominated by quantum fluctuations and are less (anti) 
self-dual. In the high temperature phase of QCD we find considerably
reduced (anti) self-duality for the modes near the edge of the spectral 
gap.
\end{abstract}
\pacs{PACS numbers: 11.15.Ha} }

Chiral symmetry breaking is one of the central pillars in our 
understanding of QCD. For the underlying mechanism an interesting picture
based on instantons
has been developed during the last 20 years \cite{instrev}. The 
relevant excitations in the QCD vacuum are believed to consist of a
weakly interacting fluid of instantons and anti-instantons. For a single
instanton or anti-instanton the Dirac operator has an exact zero-mode. 
For the weakly interacting instantons and anti-instantons instead of many zero
modes one expects the build-up of a non-vanishing density of 
small but non-zero eigenvalues corresponding to
so-called near zero-modes. This spectral density near the origin 
is related to a non-vanishing chiral condensate through the Banks-Casher 
formula \cite{BaCa80}. 

During the last year we have seen several papers \cite{Dvecs}-\cite{ivan}
which test the scenario
of chiral symmetry breaking through instantons with ab-initio
calculations on the lattice, some supporting the instanton 
picture, some challenging it. The common basic idea is to study the
properties of the near zero-modes of the lattice Dirac operator and to 
analyze whether one can identify localized structures which resemble
instantons. The underlying assumption is that the near zero-modes are only  
slight deformations of zero-modes which in turn are known to 
be localized at the same position as the instanton.

In this letter we derive a simple 
formula which relates the field strength tensor
$F_{\mu \nu}$ to a spectral sum over certain quadratic forms of the 
eigenvectors of the lattice Dirac operator. The terms in this 
spectral representation of the field strength, i.e.~the quadratic
forms of the eigenvectors, have a clear interpretation. One
can analyze the contributions of the near zero-modes 
and test the scenario of chiral symmetry breaking through instantons. 
 
In particular we will focus on analyzing the duality properties of 
the field strength. The possibility that the localized lumps in the
field strength are not (anti) self-dual, i.e.~do not have a fundamental
property of (anti) instantons, has been brought up in Ref.~\cite{horvath2},
a paper which challenges the local quantization of topological charge
in integer units. However, this possibility
is ruled out by our current results: We find that the contributions to
$F_{\mu \nu}$ which come from the near zero-modes do indeed show a 
high degree of (anti) self-duality. For contributions from 
eigenmodes with larger eigenvalues we find a decrease of (anti) 
self-duality as they are more dominated by quantum fluctuations.    

When the temperature is increased beyond its critical value, QCD
undergoes a phase transition where chiral symmetry is restored. In the 
instanton picture this transition is believed to be related to the formation 
of tightly bound instanton anti-instanton molecules. The spectrum develops
a gap and due to the Banks-Casher formula the chiral condensate vanishes. 
When analyzing the modes with eigenvalues
close to the edge of the spectral gap, we do indeed find a considerably
reduced amount of (anti) self-duality as is expected for the tightly bound
molecules.

We will denote the Dirac operator by $D = \gamma_\mu D_\mu$ 
with $D_\mu = \partial_\mu + A_\mu$. The gauge potential $A_\mu$ is 
anti-hermitian and in components is given by 
$A_\mu = - i A_{\mu}^a T_a$ where $T_a, a = 1..8$ are the 
generators of su(3). In this notation the field strength tensor
is given by $F_{\mu \nu} = [D_\mu,D_\nu]$, and in components 
reads $F_{\mu \nu} = - i F_{\mu \nu}^a T_a$.
When evaluating the square of the Dirac operator one finds
\begin{equation}
D^2(x) \; = \; \sum_\mu D_\mu^2(x) + \sum_{\mu < \nu} 
\gamma_\mu \gamma_\nu F_{\mu \nu}(x) \; .
\end{equation} 
The field strength $F_{\mu \nu}$ can be projected out by multiplying 
$D^2$ with $\gamma_\mu \gamma_\nu$ and taking the trace over the Dirac 
indices. This results in the formula ($\mu \neq \nu$)
\begin{equation}
-\frac{1}{4} \mbox{Tr}_D \Big[ \gamma_\mu \gamma_\nu D^2(x) \Big] \; 
= \; F_{\mu \nu}(x) \; ,
\end{equation}
for the field strength $F_{\mu \nu}$. For a lattice Dirac operator 
the situation is slightly more involved. On the lattice the Dirac operator
is a difference operator and has two space-time indices $x,y$. When
repeating the above calculation for e.g.~Wilson's lattice Dirac operator
$D_W$, one finds (the indices $\mu,\nu$ on the 
right hand side are not summed)
\begin{eqnarray} 
&-& \frac{1}{4} \mbox{Tr}_D \Big[ \gamma_\mu \gamma_\nu D^2_W(x,y) \Big] \; 
= \;  F_{\mu \nu}(x) \; \frac{1}{4} [ \delta_{x+\mu+\nu,y} 
\nonumber \\
&+& \delta_{x+\mu-\nu,y} + \delta_{x-\mu+\nu,y} + \delta_{x-\mu-\nu,y} ]
\; + \; {\cal O}(a^2) \; .
\end{eqnarray}
Thus one finds that the field strength is slightly smeared out 
among several lattice points. We remark that an equivalent formula holds
for staggered fermions where $\gamma_\mu \gamma_\nu$ is replaced by
a staggered factor depending on $\mu,\nu$.
For a general Dirac operator, such as the 
overlap operator \cite{overlap}, the fixed point operator \cite{perfect}
or the chirally improved operator \cite{chirimp} (which we use here) 
one finds (from now on we drop ``${\cal O}(a^2)$'')
\begin{equation}
F_{\mu \nu}(x) \; = \; - \frac{1}{4} \sum_y t_y(x)_{\mu \nu} 
\mbox{Tr}_D \Big[ \gamma_\mu \gamma_\nu D^2(x,y) \Big] \; ,
\label{traced2}
\end{equation}
with some function $t_y(x)_{\mu \nu}$ which describes the smearing out 
of the field strength around the central point $x$. For the overlap
operator this function is fast decaying in $|x-y|$ but 
has infinite support while for a
finite approximation of the fixed point operator or the chirally improved 
operator $t_y(x)_{\mu \nu}$ will vanish for a sufficiently large separation
of $x$ and $y$. One can show that a substantial 
contribution comes from $x = y$ and below we will concentrate on this 
particular case.

We now use the spectral representation of the Dirac operator to express the
right hand side of Eq.~(\ref{traced2}) in terms of the eigenvectors
$D(x,y) = \sum_j \lambda_j \; | j \rangle_x \; {}_y\langle j |$.
In this formula we denote the eigenvectors by $|j\rangle_x$ and
write explicitly only the space-time index $x$, while the color and
Dirac indices are denoted with the bra-ket notation. The corresponding 
eigenvalues are denoted by $\lambda_j$. We remark that for non-normal 
operators, the bra's ${}_y\langle j |$ have to be replaced by left
eigenvectors and only the $|j\rangle_x$ are the conventional right 
eigenvectors \cite{ivan}. 

After inserting the spectral decomposition of $D$ into our
formula (\ref{traced2}) we find:
\begin{eqnarray}
F_{\mu \nu}^a(x) & = & \sum_y t_y(x)_{\mu \nu}
\sum_j \lambda_j^2 \; f^a_{\mu \nu}(x)_{y,j} \; ,
\label{fsum1}
\\
f^a_{\mu \nu}(x)_{y,j} & = & -\frac{i}{2} \; {}_y\langle j | 
\gamma_\mu \gamma_\nu T_a | j \rangle_x  \; .
\label{fsum2}
\end{eqnarray}
In order to project onto the $a$-th component of the field strength
tensor we have used $F_{\mu \nu}^a = i 2$ Tr$_c [ F_{\mu \nu} T_a ]$.

Equations (\ref{fsum1}) and (\ref{fsum2}) provide the announced spectral
decomposition of the field strength tensor. One can now study
the individual contributions $f^a_{\mu \nu}(x)_{y,j}$ in order to analyze 
properties of the field strength tensor. It is interesting to note that the 
exact zero modes do not contribute to the spectral decomposition
(\ref{fsum1}),(\ref{fsum2}). Our formula has been tested on lattice 
discretizations of continuum instantons

Let us briefly comment on the setting of our numerical calculation. 
We study QCD in the quenched approximation using the 
L\"uscher-Weisz action  \cite{LuWeact} with coefficients from tadpole 
improved perturbation theory.
We work on lattices of size $16^4$ for the zero temperature 
ensembles and on $6 \times 20^3$ lattices for the ensembles in 
the high temperature chirally symmetric phase. The leading
coupling $\beta_1$ of the L\"uscher-Weisz action is $\beta_1=8.45$
which corresponds to a lattice spacing of $a=0.095$ fm
\cite{scale}. This value gives a temperature of 346 MeV for the 
high temperature ensemble. We use the chirally improved operator which is a 
systematic expansion of a solution of the Ginsparg-Wilson 
equation \cite{chirimp}. 
The computation of the eigenvalues and eigenvectors of the
Dirac operator was done with the implicitly restarted Arnoldi method
\cite{arnoldi}. 

\begin{figure}[t]
\centerline{\epsfig{file=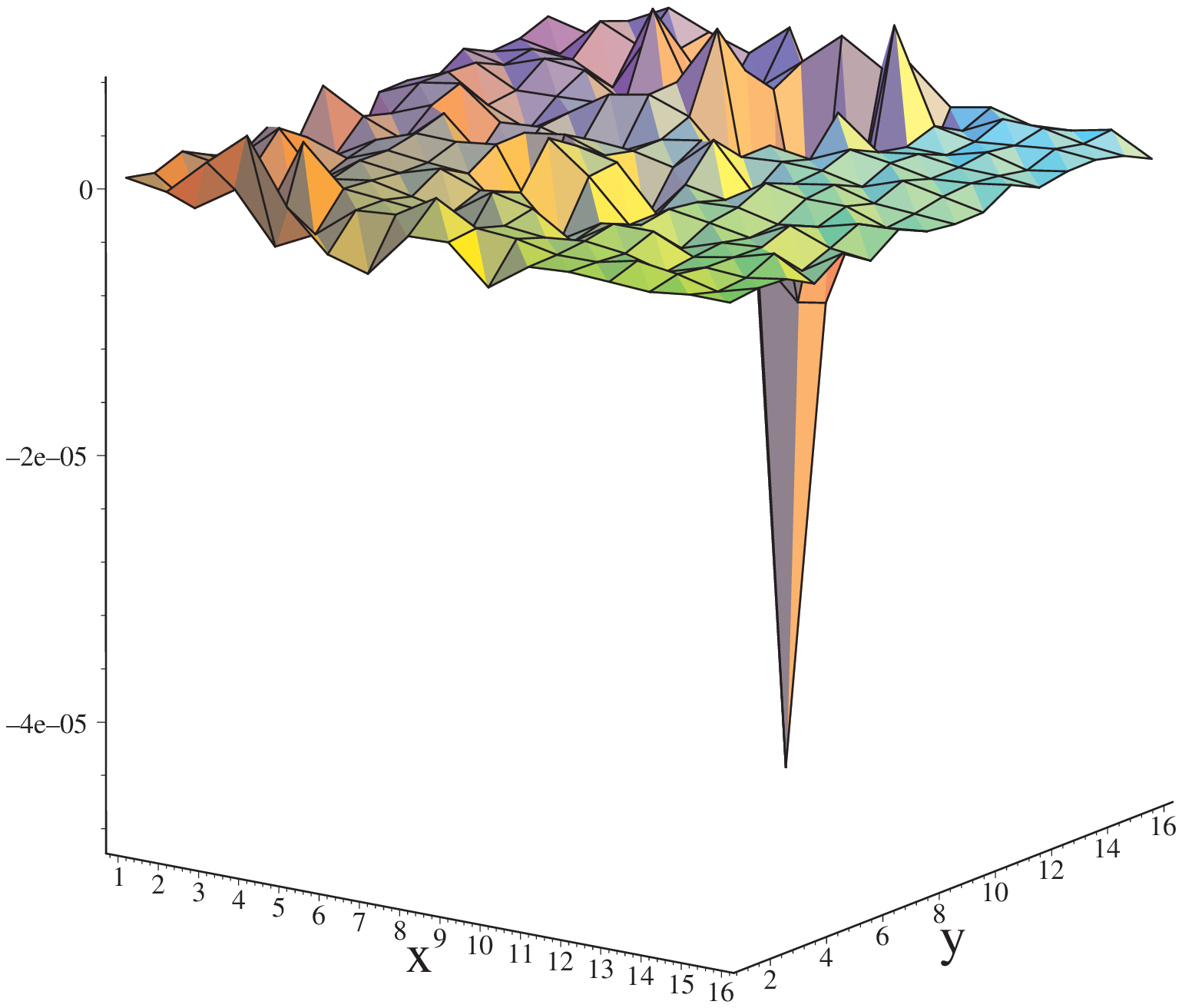,width=6.0cm}} 
\vskip5mm
\centerline{\epsfig{file=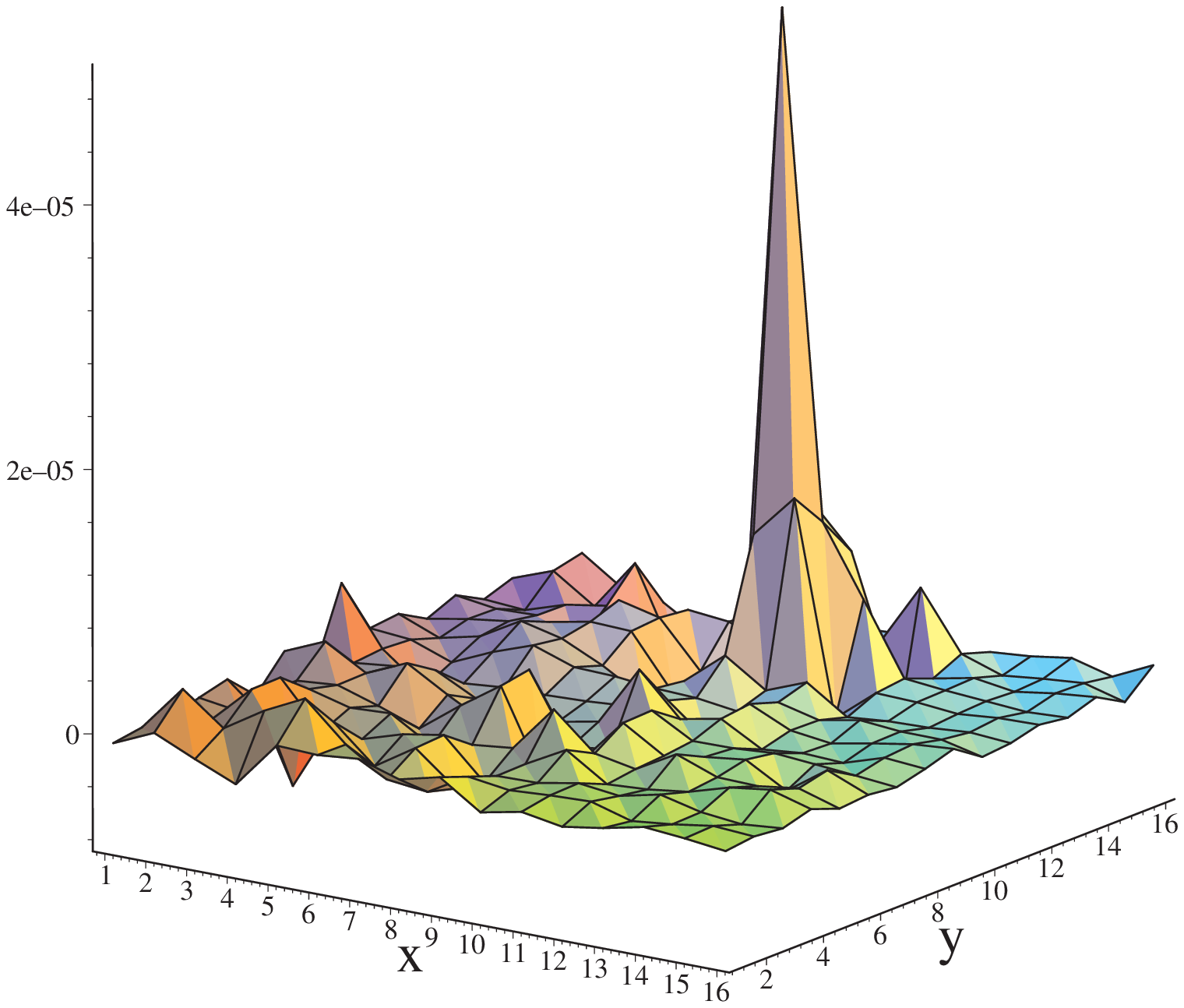,width=6.0cm}}
\vskip3mm
\caption{ Contributions $f_{12}^5$ (top plot) and $f_{34}^5$ (bottom plot)
of the first near zero-mode to the field strength. We show a slice of a $16^4$ 
lattice.
\label{fcomp}}
\end{figure}

Let us begin our analysis of the terms in the spectral representation 
(\ref{fsum1}),(\ref{fsum2}) by displaying a single contribution  
$f^a_{\mu \nu}(x)_{y,j}$ and compare it to its dual 
$\widetilde{f^a_{\mu \nu}}(x)_{y,j} = 1/2 \; \varepsilon_{\mu \nu \rho \sigma}
f^a_{\rho \sigma}(x)_{y,j}$. In particular in Fig.~\ref{fcomp} we 
show a slice through the $16^4$ lattice and display the $f_{12}^5$ component 
in the top plot and the $f_{34}^5$ component in the bottom plot. We have set 
$x=y$, i.e.~we show the center of the smeared-out field strength, and we 
evaluate the contribution for the first near zero-mode, i.e.~the eigenvector
with the smallest non-vanishing eigenvalue. 

Both plots show a large peak accompanied by smaller wiggles. The peak 
changes sign but otherwise essentially keeps its size and shape when 
going from the $f_{12}^5$ to the $f_{34}^5$ component. Thus the peak is 
anti self-dual. For the smaller wiggles it is not straightforward to
identify such a simple duality behavior - 
they are mainly quantum fluctuations. 
When inspecting many of these plots we find that the typical pattern 
consists of small quantum fluctuations and pronounced peaks which are either 
self-dual or anti self-dual. For contributions of eigenvectors with 
larger eigenvalues, so called bulk-modes, one finds that the 
abundance of isolated peaks decreases and the contributions become
more dominated by quantum fluctuations. 

\begin{figure}[t]
\centerline{\epsfig{file=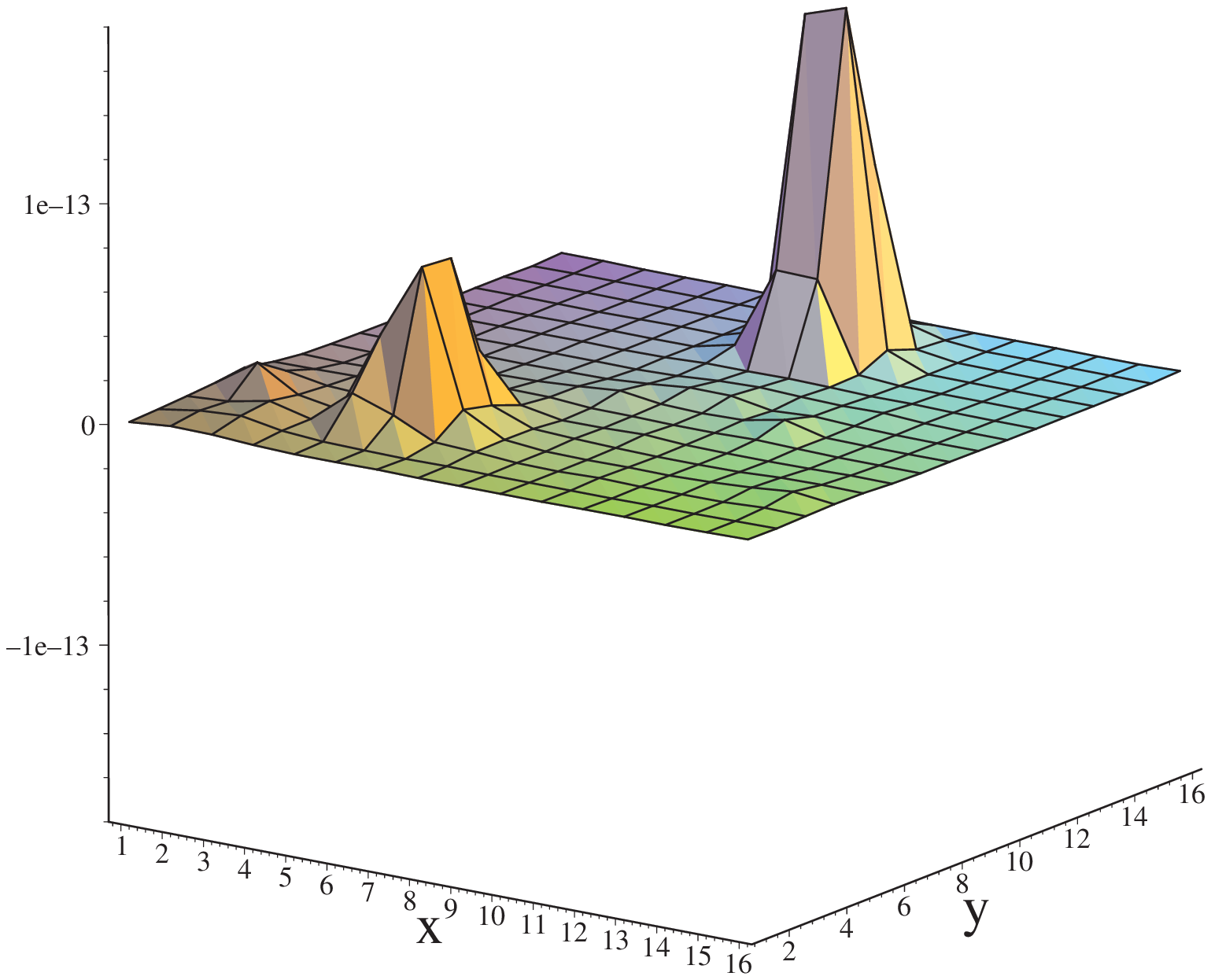,width=6.0cm}} 
\vskip1mm
\centerline{\epsfig{file=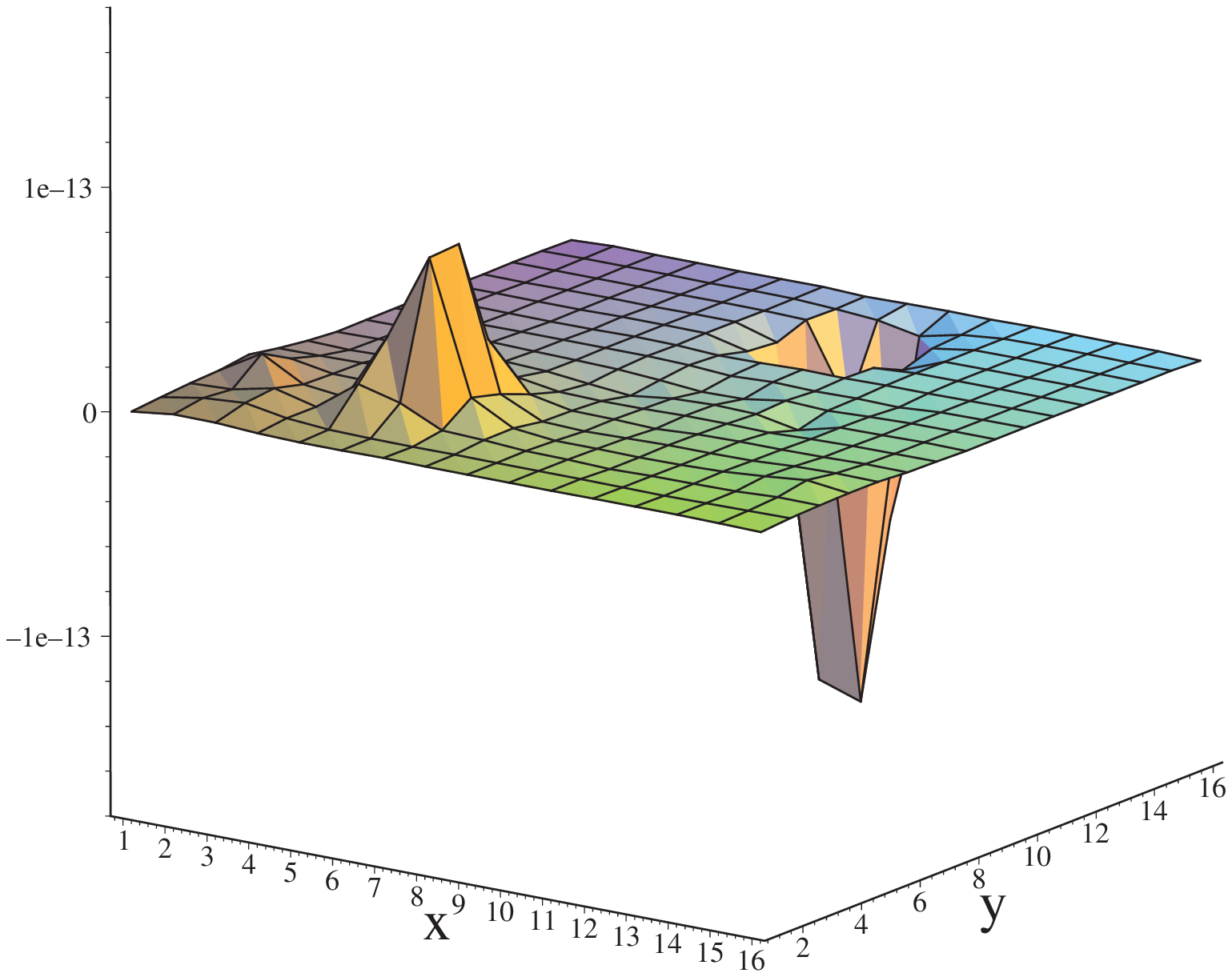,width=6.0cm}}
\vskip4mm
\caption{ Tr $F_{\mu\nu}(x)^\prime F_{\mu\nu}(x)^\prime$
(top plot) and Tr $F_{\mu\nu}(x)^\prime$ $\widetilde{F_{\mu\nu}}(x)^\prime$
(bottom plot). The primes indicate the that we take into
account only the contributions of the lowest 6 near zero-modes as defined
in Eq.(\protect{\ref{fftrunc}}).
\label{fffdual}}
\end{figure}

Quantities of more physical interest are the action density 
Tr $F_{\mu \nu} F_{\mu \nu}$ and the topological charge density 
Tr $F_{\mu \nu} \widetilde{F_{\mu \nu}}$. In particular a direct comparison 
of the two quantities shows if the field is (anti) self-dual in 
all components. Again it is most interesting 
to analyze the contributions of the near zero-modes and see whether 
they are dominated by (anti) self-dual lumps as expected from the instanton 
picture. To this purpose we truncate the sum over the eigenvalues 
$\lambda_i$ in Eq.~(\ref{fsum1}) and take into account only the 
first $N$ eigenvalues. In Fig.~\ref{fffdual} we show  
\begin{equation}
\mbox{Tr} \, F_{\mu \nu}(x)^\prime F_{\mu \nu}(x)^\prime  =  
\sum_{i,j = 1}^N \frac{\lambda_i^2 \lambda_j^2}{2} f^a_{\mu \nu}(x)_{x,i}
f^a_{\mu \nu}(x)_{x,j} ,
\label{fftrunc}
\end{equation}
in the top plot. The color indices $a = 1 .. 8$ on the right hand side 
of (\ref{fftrunc}) are summed to produce the color trace on the left hand side.
The corresponding contribution to 
Tr $F_{\mu \nu}(x)^\prime \widetilde{F_{\mu \nu}}(x)^\prime$ is 
constructed in the same way and is displayed in the  
bottom plot of Fig.~\ref{fffdual}. For the figure we set $N=6$, 
i.e.~we take into account only the contributions from the lowest 6 
near zero-modes. 

Both plots in Fig.~\ref{fffdual} show two pronounced peaks with one
of them changing sign when going from 
Tr $F_{\mu\nu}(x)^\prime F_{\mu\nu}(x)^\prime$ to 
Tr $F_{\mu\nu}(x)^\prime \widetilde{F_{\mu\nu}}(x)^\prime$. We remark
that Figs.~\ref{fcomp} and \ref{fffdual} were made from the 
same gauge configuration and the anti self-dual sign-changing peak 
of Fig.~\ref{fffdual} is the same peak which we already saw in the 
contribution to the 5-component displayed in Fig.~\ref{fcomp}. The other
smaller peak in Fig.~\ref{fffdual}
is a self-dual fluctuation in a different color component.  
Since the composed quantities Tr $F_{\mu\nu}(x)^\prime F_{\mu\nu}(x)^\prime$
and Tr $F_{\mu\nu}(x)^\prime \widetilde{F_{\mu\nu}}(x)^\prime$ are 
built from products of the single components $f_{\mu \nu}^a$ the
relative size of the quantum fluctuations is suppressed considerably
compared to the larger, (anti) self-dual structures. We find that 
the contributions of the near zero-modes to 
Tr $F_{\mu\nu}(x)^\prime F_{\mu\nu}(x)^\prime$ and 
Tr $F_{\mu\nu}(x)^\prime \widetilde{F_{\mu\nu}}(x)^\prime$
are entirely dominated by lumps which are either self-dual
or anti self-dual. 

\begin{figure}[t]
\centerline{\epsfig{file=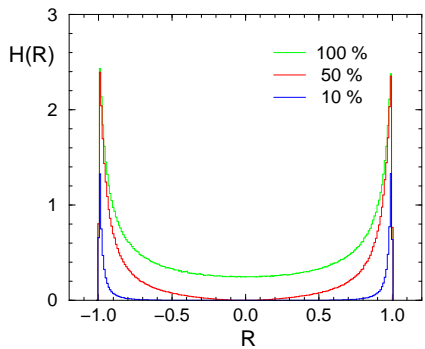,width=5.5cm}}
\caption{Histograms for the observable $R$ which tests the 
(anti) self-duality of the near zero-modes 
(compare Eq.~(\ref{selftest1})). 
We display our data without
a cut (top curve), a cut of 50\% (middle curve) and a cut of 10 \%
on the number of lattice points supporting the highest peaks of 
$|$Tr$F_{\mu \nu}(x)^\prime \widetilde{F_{\mu \nu}}(x)^\prime|$. 
\label{selfhistoz}}
\end{figure}

In order to go beyond an illustration of the duality properties
by a few examples we now discuss an observable which allows
to test (anti) self-duality systematically. Its construction is similar
to the chirality observable proposed in \cite{horvath1}. 
We define the ratio 
\begin{equation}
r(x) \; = \; \frac{
\mbox{Tr} \, F_{\mu\nu}(x)^\prime F_{\mu\nu}(x)^\prime \, - \, 
\mbox{Tr} \, F_{\mu\nu}(x)^\prime \widetilde{F_{\mu\nu}}(x)^\prime }{
\mbox{Tr} \, F_{\mu\nu}(x)^\prime F_{\mu\nu}(x)^\prime \, + \, 
\mbox{Tr} \, F_{\mu\nu}(x)^\prime \widetilde{F_{\mu\nu}}(x)^\prime } \; .
\label{selftest1}
\end{equation}
For a space-time point $x$ where the gauge field is self-dual
the numerator will vanish while the denominator is finite and $r(x)$ 
equals to 0. Conversely for an $x$ where the gauge field is anti self-dual 
the role of numerator and denominator are exchanged and $r(x) = \infty$.
For space-time points without definite duality properties $r(x)$ assumes
some finite value between 0 and $\infty$. The transformation 
$R(x) = 4/\pi \arctan r(x) - 1 $
maps the interval $[0,\infty)$ into the interval $[-1,1]$. For
configurations which are dominated by (anti) self-dual lumps
one expects values near $\pm 1$. As for the local chirality variable of 
\cite{horvath1} it is interesting to study different selections for
the lattice points $x$ in $R(x)$. 
In Fig.~\ref{selfhistoz} we use all lattice points (top curve),    
the subset of 50\% of the lattice points supporting the highest peaks of
$|$Tr $F_{\mu \nu}(x)^\prime \widetilde{F_{\mu \nu}(x)^\prime}|$ 
(middle curve) and also a cut of 10\% (bottom curve). Again we use the
6 lowest modes in the series for $F_{\mu \nu}(x)^\prime$.
The histograms were computed by averaging over 50 configurations. 

Fig.~\ref{selfhistoz} shows that the contributions of the near zero-modes
to the spectral representation of the field strength are highly 
(anti) self-dual. Even when including all lattice points (100\%, top curve) 
one finds a pronounced double peak. When throwing away 50\% of the
lattice points with small 
$|$Tr$F_{\mu \nu}(x)^\prime \widetilde{F_{\mu \nu}}(x)^\prime|$
this cuts mainly into the center of the histogram, showing that 
the most (anti) self-dual excitations are 
indeed the large peaks sticking out of the quantum fluctuations.
This trend continues when 
focusing on only the highest 10\% of the peaks. Since our analysis is
based on the exact spectral decomposition of the field strength it truly 
reflects self-duality of the topological lumps in the gauge field.
\begin{figure}[t]
\centerline{\epsfig{file=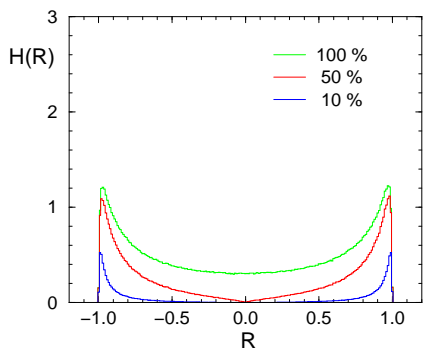,width=5.5cm}}
\caption{The same as in Fig.~\ref{selfhistoz} but now for the
ensemble in the chirally symmetric phase of QCD. 
\label{selfhistot}}
\end{figure}

It is interesting to perform the same analysis also in the chirally
symmetric phase of QCD. As discussed, the instanton picture describes 
this phase by tightly bound 
molecules of instantons and anti-instantons \cite{instrev}.
In Fig.~\ref{selfhistot} we show our results for the histograms of
the duality observable $R$ for our ensembles in the high temperature phase. 
Again we average over 50 configurations. $F_{\mu \nu}(x)^\prime$ is now
computed from the 6 modes with eigenvalues closest to the edge of
the spectral gap. Obviously the amount of (anti) self-duality has decreased 
considerably which is exactly what is expected to happen when instantons
and anti-instantons condense to tightly bound molecules and
lose their identity as isolated (anti) self-dual objects.

Let us briefly summarize the obtained results.
We derive a representation of the field strength in terms 
of a spectral sum of quadratic forms of the eigenvectors of the 
lattice Dirac operator. The contributions of individual eigenvectors to
the field strength can be analyzed. We find that the near zero-modes
are dominated by (anti) self-dual lumps, while modes with larger eigenvalues 
further in the bulk of the spectrum become dominated by quantum fluctuations. 
Our spectral representation thus decomposes the field strength tensor
into contributions (the near zero-modes) sensitive to the large 
scale fluctuations and contributions (the bulk modes) dominated by
quantum fluctuations. In the high temperature phase we find 
a reduced amount of (anti) self-duality for the modes with eigenvalues 
near the edge of the spectral gap. Our results support the picture of
chiral symmetry breaking by instantons and its restoration through
the formation of instanton anti-instanton molecules.  
\vskip3mm
\noindent
{\bf Acknowldegements:} 
I would like to thank Pierre van Baal, Tom DeGrand, Meinulf G\"ockeler, 
Ivan Hip, Christian Lang, Paul Rakow, Stefan Schaefer and Andreas Sch\"afer
for discussions and the Austrian Academy of Science for support 
(APART 654). The calculations were done on the Hitachi SR8000
at the Leibniz Rechenzentrum in Munich and I thank the LRZ staff for
training and support.

\end{document}